\documentstyle[12pt,aasms4]{article}
\begin{document}

\title{$HST$ Imaging Polarimetry of the Gravitational Lens  
$FSC10214+4724$\altaffilmark{1}}
\author{Hien T. Nguyen\altaffilmark{2},
Peter R. Eisenhardt\altaffilmark{2},
Michael W. Werner\altaffilmark{2},
Robert Goodrich\altaffilmark{3},
David W. Hogg\altaffilmark{4,5},
Lee Armus\altaffilmark{6},
B. T. Soifer\altaffilmark{7}, 
G. Neugebauer\altaffilmark{7} }

\authoremail{hien@cougar.jpl.nasa.gov}

\altaffiltext{1}{Accepted for publication in Astronomical Journal,
February 1999.  Based on observations made with NASA/ESA $Hubble$
$Space$ $Telescope$, obtained at the Space Telescope Science Institute, which
is operated by AURA, Inc., under NASA contract NAS 5-26555.}

\altaffiltext{2}{MS 169-506, Jet Propulsion Laboratory, California
Institute of Technology, Pasadena, CA 91109}

\altaffiltext{3}{California Association for Research in Astronomy,
65-1120 Mamalahoa Highway, Kamuela, HI 96743}

\altaffiltext{4}{Institute for Advanced Study, Olden Lane, 
Princeton, NJ 08540}

\altaffiltext{5}{Hubble Fellow}

\altaffiltext{6}{SIRTF Science Center, California Institute of Technology,
Pasadena, CA 91125}

\altaffiltext{7}{Division of Physics, Math and Astronomy, California
Institute of Technology, Pasadena, CA 91125}

\begin{abstract}

We present imaging polarimetry of the extremely luminous, 
redshift 2.3 $IRAS$ source $FSC10214+4724$.  The observations were obtained 
with $HST$'s Faint Object Camera in the F437M filter, which is free of 
strong emission lines.  The 0.7 arcsec long arc is unresolved to 0.04 arcsec
FWHM in the transverse direction, and has an integrated 
polarization of $28 \pm 3$ percent, in good 
agreement with ground-based observations.  The polarization position angle 
varies along the arc by up to $35$ deg.
The overall position angle is $62 \pm 3$ deg east of north.
No counterimage is detected to $B = 27.5$ mag ($3\sigma$), giving an
observed arc to counterimage flux ratio greater than 250, considerably
greater than the flux ratio of 100 measured previously in the $I$-band.
This implies that the configuration of the object in the source plane at the 
$B$-band is different from that at $I$-band, and/or that the lensing 
galaxy is dusty.

\end{abstract}


\section{Introduction}

With a redshift of $z = 2.286$, $FSC10214+4724$ remains the most 
distant object identified in the $IRAS$ database.  At the time of its 
initial discovery in 1991 (Rowan-Robinson {\it et al.})\markcite{RR91}, 
it was estimated 
to have a luminosity well in excess of $\sim 10^{14} L_{\sun}$, most
of which emerged in the far-infrared.   Subsequent $HST$ observations of 
$FSC10214+4724$\markcite{E96} (Eisenhardt et al. 1996, hereafter E96) 
with F814W (roughly $I$-band, center wavelength of 814nm) have shown 
conclusively that 
the source is gravitationally lensed by a foreground elliptical galaxy, 
reducing its intrinsic luminosity to $\sim 2 \times 10^{13} L_{\sun}$.  
The redshift of the lensing galaxy has 
not yet been measured with certainty, but a value of 0.9 is 
favored by E96 and Lacy, Rawlings $\&$ Serjeant (1998).
Millimeter and sub-millimeter line and continuum 
observations have shown that $FSC10214+4724$ contains huge amounts of 
interstellar dust and gas and have raised the possibility that it is 
powered by a massive starburst (Downes, Solomon $\&$ Radford 1995;
Rowan-Robinson et al. 1993; Scoville et al. 1995).  Visual
spectropolarimetry\markcite{G96} (Goodrich et al. 1996, hereafter
G96) has revealed both a highly polarized (25 percent at 
0.44 $\mu$m) continuum and polarized broad wings on the otherwise 
unpolarized emission lines.  This is the polarimetric signature of a 
dust-embedded quasar, hidden from direct view but seen in 
reflection via scattering off of favorably placed cloud(s).
Therefore both stellar and non-thermal energy sources probably
contribute to the luminosity of this object\markcite{K96} 
(Kroker et al. 1996, Lacy, Rawlings $\&$ Serjeant). 

The present program makes use of the intevening
gravitational lens as a microscope to study the structure 
of the inner regions of the quasar.  At the F814W bandpass, 
the magnification appears to be
quite high ($\sim 100$, E96) suggesting that the 
background source is extremely close to a caustic of the lensing 
potential, so that small changes in the source position or size 
with wavelength will lead to differences in the position and 
structure of the image which could be resolved with $HST$.  The
principal features of the lensed image at F814W are a 
0.7 arcsec long arc which is unresolved ($< 0.06$ arcsec) in the 
transverse direction, and an unresolved counterimage, about 1.6 arcsec 
away, with about 1 percent of the total flux of the arc.  The absolute 
and relative positions of the arc and the counterimage, the extent and 
structure of the arc, and 
the relative brightness of the arc and the counterimage, are all 
sensitive to the structure of the source at the observed wavelength 
and its position relative to the caustic of the lensing potential (E96).

This paper reports the results of imaging polarimetry of 
$FSC10214+4724$, carried out in the continuum using filter
F437M (roughly $B$-band, center
wavelength of 437nm) with the $HST$'s Faint Object Camera (FOC).  
Subsequent papers will report the results 
of $HST$ imaging of the source in narrow bands centered on a number 
of emission lines, providing information on the spatial distribution 
of gas in  the narrow line region of the quasar.

At the $FSC10214+4724$ redshift of $z = 2.286$, one 0.014 arcsec FOC pixel 
subtends 100(60) h$^{-1}$ pc for $q = 0(0.5)$, where $h=H_o/100$ km s$^{-1}$ 
Mpc$^{-1}$ (where not otherwise specified, we assume 
$H_o=50$ km s$^{-1}$ Mpc$^{-1}$ and $q=0.5$).  The FWHM of the 
$HST$'s point spread function at F437M is $0.042$ arcsec.
Hereafter, F437M and F814W are
interchangeably referred to as $B$ and $I$-band, respectively.

\section{Observations and Reduction}

%\subsection{Observations}

Imaging polarimetry was performed through three separate polarizers,
referred to as POL0, POL60 and POL120, where the number indicates the
position angle with respect to the detector in an instrumental coordinate 
system.  A total of 15 orbits (35,055 seconds) of polarimetry data were
obtained using the F437M filter in the FOC, with 0.014 arsec per pixel and
the field of view is 7.16$\times$7.16 arcsec$^2$.  The data were obtained 
in three visits.  In the first visit, one orbit with each polarizer was
obtained on 19 May, 1997.  Each orbit began with a 200 second exposure 
on the $B$=19 mag star 13.2 arcsec to the east (star A in E96), to
allow registration of data in the three polarizers and to check the PSF;
the remainder of each orbit was spent on the arc.  In the second visit, six
orbits were obtained on 20-21 May 1997, with each orbit evenly divided
between the three polarizers. The position angle orientations on the
sky for the second visit were the same as for the first.  In the third
visit, six orbits were obtained on 29 June 1997 using the same strategy 
as in the second visit, except that the telescope was rotated around its 
line of sight so that the polarizer position angles were rotated 30 deg 
counterclockwise on the sky from first and second visit.

%\subsection{Reduction}

After standard processing provided by STScI, the FOC frames were 
corrected for spatial shifts introduced by the different polarizers.  
The shifts 
($\sim$ few pixels) were determined by measuring the position
of star A relative to each polarization image.  Additional 
corrections were also applied for dithers (visit 2 and 3) and 
rotation (visit 3).  The background for each individual frame was
estimated and subsequently subtracted.  The resulting frames were
corrected for calibration and co-added for each visit.  
Following Thompson $\&$ Robinson (1995)\markcite{TR1995}, 
we define the total intensity,
\begin{equation}
I = \frac{2(s_0 + s_{60} + s_{120})}{3}
\end{equation}
where $s_i$ is the flux through each polarizer.
The normalized Stokes parameters are then given by,
\begin{equation}
q = \frac{Q}{I} = \frac{2s_0 - s_{60} - s_{120}}{s_0 + s_{60} + s_{120}}
\end{equation}
\begin{equation}
u = \frac{U}{I} = \frac{\sqrt{3}(s_{60} - s_{120})}{s_0 + s_{60} + s_{120}}
\end{equation}
and the fractional polarization, 
\begin{equation}
p = \sqrt{u^2 + q^2}
\end{equation}
and the polarization position angle, which is measured counterclockwise
from the scan direction ($x$-axis),
\begin{equation}
\theta = \frac{1}{2}arctan\frac{u}{q} + \theta_0
\end{equation}
where $\theta_0$ is the zero point offset of the polarizers' orientation
with respect to the detecor scan direction and is given to be
$-1.4^o$.

The overall polarization results were then obtained by averaging the 
Stokes parameters of the three visits.  
 
\section{Results}

The F437M image of $FSC10214+4724$ is shown in Figure 1, 
in grayscale, obtained from a straight co-add of frames from filters 
POL0 and POL120 for all three visits.  
These POL0 and POL120 data had a diffraction limited
PSF with a FWHM of 0.042 arcsec. 
The POL60 data had a degraded PSF, and thus 
were left out for this intensity map.
However, in the polarimetry analysis where POL60 images were included, 
and the pixels were binned to exchange higher spatial 
resolution for improvements in the signal-to-noise ratio, 
this degradation does not significantly affect the results. 

\subsection{Morphology}

In the F437M FOC data, the $FSC10214+4724$ arc has a length of
$\sim 0.7$ arcsec roughly along the east-west direction and is 
essentially unresolved ($< 0.04$ arcsec) in the transverse direction.  
This is the same basic morphology seen by E96 in the F814W
WFPC2 data.  The tangential profile exhibits significant structure.  
Summed over 3$\times$3 pixels to improve the signal-to-noise ratio, the arc 
is seen to have two prominent peaks.  The centroids of these two 
peaks are separated by at least 0.30 arcsec.  This is significantly larger 
than the 0.24 arcsec separation observed in the F814W image.

The four objects seen in the F814W image which fall in the FOC 
field-of-view are the arc (refered to as component 1 by E96), 
the two lensing galaxies (components 2 and 3) and the counterimage 
(component 5).  Components 1, 2 and 5 are shown as contours and labelled
in Figure 1.  The only object visible 
in the FOC data is the arc, as seen in grayscale in Figure 1.  
The absence of components 2 and 3  
in the FOC data is expected from their measured spectral energy 
distribution (E96) and the depth of the FOC image (hereafter, 
E96's component 2 is referred to as the lensing galaxy).  
Most significantly, the counterimage seen in the F814W image and in 
the K band (Graham $\&$ Liu 1995,\markcite{GL}, Broadhurst $\&$ Leh\'ar 
1995\markcite{BL}) is not 
present in the F437M FOC data.  The ratio of the flux of the arc to 
the $3\sigma$ level of the background noise at the expected position of 
the counterimage is 250.  This limit is 2.5 times greater than the 
arc-to-counterimage flux ratio ($\sim 100$) in the F814W image (E96). 

Our coadded image gives an integrated flux density over the length
of the arc of 1.6$\times 10^{-17}$ ergs cm$^{-2}$ s$^{-1}$ ${\AA}^{-1}$,
equivalent to $B = 21.5$ mag.  This is similar to the integrated magnitude 
obtained by Broadhurst $\&$ Leh\'ar.

\subsection{Astrometry}

Astrometry provides critical information needed for mapping the image 
into the source plane.  The task is complicated due to the fact that 
there is no object other than the arc itself in the FOC field of view.  
The accuracy of the measurement thus depends upon the 
knowledge of the distance from the reference object, star A, to the 
lensing galaxy, and the precision offsetting of $HST$.  From the F814W data,
Star A was measured to be $13.19 \pm 0.01$ arcsec west and 
$0.99 \pm 0.02$ arcsec north of the lensing galaxy.  The errors 
are due to uncertainties in the correction for geometrical distortion and 
in the determination of the centroid
of star A, whose core is saturated in the F814W PC data.  
Assuming the telescope pointing between the FOC image of star A and the arc
is accurately known, our best determination of the relative 
positions of the arc at F437M and F814W is shown in Fig 1.  
Relative to the $I$-band arc (in contours), the arc as seen by 
the FOC data (in grayscale) appears to be shifted radially towards the 
lensing galaxy, by $12 \pm 14$ mas.  Here we have root-sum-squared
errors of 10 mas the for geometrical distortion and centroiding, and
10 mas for telescope pointing (R. Jedrzejewski, private communiation).

%\subsection{Photometry}

\subsection{Polarization}

Our polarization measurement confirms that $FSC10214+4724$ 
is highly polarized.  The integrated linear polarization of 
the entire arc, determined from the total fluxes measured
with each of the three polarizers, is $31 \pm 3$ percent for the first
visit, $24 \pm 5$ percent for the second and $26 \pm 5$ percent
for the third.  The average degree of polarization then
is $28 \pm 3$ percent, consistent with the ground-based observations 
\markcite{G96}(G96).  The position angles, measured east of north, 
for the three separate visits appear to be in satisfactory agreement.  
They are $59 \pm 3.5$ deg for the first visit, $61 \pm 7$ deg for 
the second and $67 \pm 5$ deg for the third.  The overall orientation, after 
averaging the Stokes parameters from the three visits, is $62 \pm 2.7$ deg, 
compared to the ground-based observations of $69.9 \pm 0.2$ deg by G96, 
and $75 \pm 3$ deg by Lawrence et al. (1993)\markcite{AL93}.  Lawerence et al. 
integrated their polarimetry data over the 400 to 1000 nm range.  G96
obtained spectropolarimetry over the same spectral range as Lawrence et al., 
and their position angle appeared to be constant over that range.
The discrepancy in the position angles is not understood,
but may be due to inclusion of low surface brightness regions in the
ground-based data which are not detected by $HST$.

Taking advantage of the unresolved nature of the arc in the transverse 
direction, a high resolution one-dimensional polarization map was produced 
by summing over $10{\times}3$ pixels, in the transverse and tangential 
directions, respectively.  The choice of number of pixels was made so 
as to improve the signal-to-noise ratio while preserving the structure 
along the tangential direction.  The combination of these pixels gives 
substantial improvement in signal-to-noise ratio and minimizes the effect 
of misregistration and differing PSF, thus allowing reliable determination 
of the polarization for each combined pixel. As shown in Figure 2, the 
total intensity I, the degree of polarization P and the position angles 
vary significantly along the arc.  The maximum change in the position 
angle, $\Delta\theta_{max}$, is $35 \pm 5$ deg.

\section{Discussion}

There are no sources other than $FSC10214+4724$ detected in the FOC
field of view, even though the image is very deep, with a $3\sigma$ point 
source detection limit of $B\approx 27.5$~mag in the 
central $6\times 6~{\rm arcsec^2}$.  This is consistent with
the expectation from faint field galaxy source counts (Williams et al. 
1996\markcite{REW}).  

\subsection{Absence of the counterimage}

The absence of the counterimage (component 5 in E96) in the $B-$band FOC 
image is surprising.  This may be due either to larger lensing
magnification in the F437M bandpass than in F814W seen by E96, or 
to extinction by dust in the lensing galaxy.  We address each of 
these possibilities in turn.  

Although gravitational lensing is achromatic,
the magnification is a sensitive function of source size and position 
near the caustic in the source plane.
Different distributions for the UV and optical continuum
regions can account for the appearances of $FSC10214+4724$ at 
different wavelengths noted by Matthews et al. (1994).  
The lens model suggested by E96 and a source geometry 
as sketched in Figure 3 reproduce qualitatively the inferred lensing 
magnification of 250, the observed arc morphology and the arc astrometry 
at F437M.  Our source model puts the $B$-band 
source with a radius of $\sim 20$ pc, right on the cusp in the caustic line, 
approximately 100 pc away from the center of the $I$-band source.
The fact that the $B$-band source lies on the caustic is not 
entirely fortuitous but perhaps due to "magnification bias".
In reality the scattering clouds may be widely distributed, but only the 
regions near the cusp are highly magnified and readily detected.  
Because of uncertainties in telescope pointing, we take the observed
difference in angular position of the arc to be an upper limit rather
than a detection.  However, the small differences in morphology of the
arc, when combined with the lens model do indeed suggest a small
offset between the $B$-band and $I$-band sources, and hence a small offset
between the $B$-band and $I$-band arcs of roughly 10 mas, in the direction
provided by the telescope astrometry (section 3.2).  For the simple
quasar picture in which a scattering cloud is illuminated by a 
hidden AGN in the source plane, the positional offset of the $B$-band 
cloud relative to that in the $I$-band suggests that the direction 
from the hidden AGN to these sources could be different, and perhaps 
could give rise to different position angles (since the polarization vector is 
perpendicular to the direction to the AGN).  However, existing 
spectropolarimetric data indicate that position angle is independent of 
wavelength (Goodrich, private communication).

If in fact there is no radial shift between the $I$ 
and $B$ band data sets, reddening in the lensing galaxy may still explain
the absence of the counterimage at $B$ band, since the counterimage
and the arc are viewed through different parts of the elliptical lensing
galaxy, and therefore may be affected by different amounts of extinction.
Dusty gravitational lenses have been observed (e.g., MG 1131+0534 
Larkin et al. 1994 and MG 0414+0534 Lawrence et al. 1995).  If it is
assumed that the arc is not significantly reddened since its optical
path is $\sim 10$ kpc away from the center of the lens galaxy, $A_V > 0.67$ 
mag at $z = 0.9$ is required to produce $E(B - I) > 1$ mag observed 
for the counterimage, using the extinction law from Cardelli, Clayton, $\&$ 
Mathis 1989 ($R_V = A_V/E(B-V) = 3.1$).  This optical depth is similiar 
to that found in some local elliptical galaxies (Goudfrooij \& de~Jong 1995).  
This extinction, if present, would reduce the intrinsic lensing 
magnification ratio estimated from the F814W data to less than 45, 
which is closer to the magnification ratio of roughly 30 estimated 
in the far infrared (E96).  The source structure and hence the polarization 
angle could then be very similar for both F437M and F814W. 

\subsection{Pattern of Position Angles}

Given a lensing magnification, it is possible to infer the source diameter,
$D$, using the lens model (E96).  The observed pattern of position angles 
then may be directly translated into the source plane, and can be used to
infer the location of the dust enshrouded AGN.  For the quasar model above, 
the projected distance, $R$, from the continuum region to the hidden AGN 
may then be estimated by the simple relation, 
$D \approx R\Delta\theta_{max}$, where $\Delta\theta_{max}$ is the 
maximum change of the position angle.  For a range of lensing magnification 
of 250 to 45, $D$ varies from 40 to 100 pc, 
so $R$ is less than 160 pc (see also, E96).  
This smale scale is generally inaccessible at these wavelengths for 
objects at redshifts $z > 2$.  Notice this range of $R$ is similar
to the distance from the active nucleus to the continuum region found for 
the local Seyfert 2 galaxy, NGC1068 (Capetti et al. 1995).  

The relative position of the source and the caustic is crucial in 
explaining the many features of the arc.  For example, if some part of 
the source is within the cusp, that part will be triply imaged in the arc
such that the parity of the first image is opposite to the parity of the 
second, and is the same as that of the third.  A monotonic position-angle 
variation across the source located within the cusp will naturally 
produce the rising, falling and rising position angle pattern shown 
in the central part of Figure 2c. 

\subsection{Future Observations}

Either change in $FSC10214+4724$ source geometry 
or extinction in the lensing galaxy could
be responsible for the fact that the counterimage is present in 
the $I$-band, WFPC2 data while absent in the $B$-band, FOC data.
We expect much of this ambiguity to be resolved by a more accurate 
determination of the relative astrometry of the $FSC10214+4724$ 
system with planned $HST$ WFPC2 observations in the F467M and F814W.
In the present study, the morphological differences between the
unpolarized $I$-band and polarized $B$-band images of $FSC10214+4724$ are
not large, because the differences in size and location of the regions
emitting these components are small relative to the resolution of the
high-magnification gravitational lens combined with HST.
However the same is not expected to hold for the narrow emission line
region (the NLR).  From detailed studies of AGN at low redshifts,
it is known that the NLRs and the extended NLRs can exhibit 
complex morphologies over hundreds of parsecs to kiloparsec scales 
(e.g. Pogge 1989, Wilson $\&$ Tsvetanov 1994).  At high redshifts, the 
emission-line nebula around powerful AGN can be quite spectacular, 
reaching sizes of 10-100 kpc (e.g., McCarthy, Spinrad
$\&$ van Breugel 1995, Armus et al. 1998).  If the NLR of 
$FSC10214+4724$ has similar properties, its appearance through filters 
isolating specific UV and optical emission lines should be quite 
different from its broad band, continuum morphology.  For example, 
if the NLR is displaced towards the lensing galaxy, the arc will split up 
into three images and the counterimage will become relatively more
prominent.  If displaced away, the arc will become shorter and the
counterimage will again become relatively more prominent.  The displacement, 
shape and size of the NLR may be determined from the observed morphology and
the lens model.  This makes $FSC10214+4724$ a unique system for direct
imaging of quasar emission regions at high redshift.  Observations
are planned in $HST$ cycles 7 and 8 to image this system in
narrow-band filters centered on the narrow emission lines, 
[C IV]$\lambda$1549, [Ne V]$\lambda$3426 and [O III]$\lambda$5007.

\section{Summary and Conclusions}

We have reported imaging polarimetry of the
gravitationally-lensed object $FSC10214+4724$, carried out with 0.04 arcsec
resolution at F437M (roughly $B$-band) using the Faint Object 
Camera on $HST$.  The principal results of this work are the following:

a.  The F437M source appears as an unresolved arc, 0.7 by $ < 0.04$ arcsec,
coincident to within the errors with the arc seen at F814W by E96.

b.  The absence of a counterimage suggests that the apparent magnification
at F437M is 250, considerably higher than measured at F814W and inferred
for the far infrared (E96).  It is possible that this
effect is due to differential extinction in the lensing galaxy.  However,
if further work shows that that the magnification is in fact higher at
F437M than at F814W, then the emitting region at the shorter wavelength
must be smaller and closer to the caustic of the lensing potential.  The
size and position difference might be a a consequence of patchy UV
extinction in the high luminosity core of $FSC10214+4724$ (Lacy, Rawlings
$\&$ Serjeant).

c.  The net polarization of the arc at F437M is consistent with what has
been determined from ground-based measurements, but we determine that the
position angle of polarization rotates systematically with position along
the arc.  The total variation of the angle is about 35 deg.  Under the
assumption that the polarization is due to scattering of continuum
radiation from an embedded source, the projected distance of the scattering
region from this source is less than 160 pc; it could be as small as 60 pc if
the magnification is in fact as high as 250.

\acknowledgments

We thank Roger Blandford and Roger Hildebrand for helpful discussions 
and Robin Evans for his help with corrections for geometrical distortion 
of the PC data.  HTN wishes to thank the Institute for Advanced Study 
for its hospitality during his visit for a part of this work. 
Support for DWH was provided by Hubble Fellowship grant
HF-01093.01-97A from STScI, which is operated by AURA under NASA
contract NASA5-26555.
Support for this work was provided by NASA through grant number
GO-6834.01-95A from STScI, which is operated by AURA under NASA 
contract NAS5-26555.
Portions of the research described in this paper were carried out
by the Jet Propulsion Laboratory, California Institute of Technology,
under a contract with NASA.

\clearpage

\figcaption[fig1.ps]{
$HST$ Faint Object Camera image of $FSC10214+4724$ in filter F437M in
grayscale.  The overlayed contours are from the PC F814W data (E96),
with the centroid of the lensing galaxy and counterimage labelled
as 2 and 5, respectively.  Notice the absence of the lensing galaxy 
and the counterimage in the F437M FOC data.  North is $76.0$ deg 
clockwise from vertical.}

\figcaption[fig2.ps]{One dimensional profiles of $FSC10214+4724$ in
filter F437M.  Total intensity I (panel a), Degree of polarization P 
(panel b) and position angle $\theta$ (panel c) vs. position in arcsec 
along the arc.  The smoothing box for the data was 3 pixels parallel
to the arc $\times$ 10 pixels transverse to the arc.  East is to the right.}

\figcaption[fig3.ps]{
A source model of $FSC10214+4724$ for lensing magnification of $\sim 250$. 
Solid lines: Position and size of the source in $I$ and $B$ bands.  Solid 
vector is the position angle measured by this work.  Dotted vectors are 
the position angle measured by G96.  Broken line: the cautic.  Dashed line: 
the location of hidden AGN inferred from the variation of the position 
angles of polarization observed along $FSC10214+4724$ arc.  (0,0) is the
position of the lensing galaxy in the source plane.  North is 37.1 deg 
counterclockwise from vertical.}

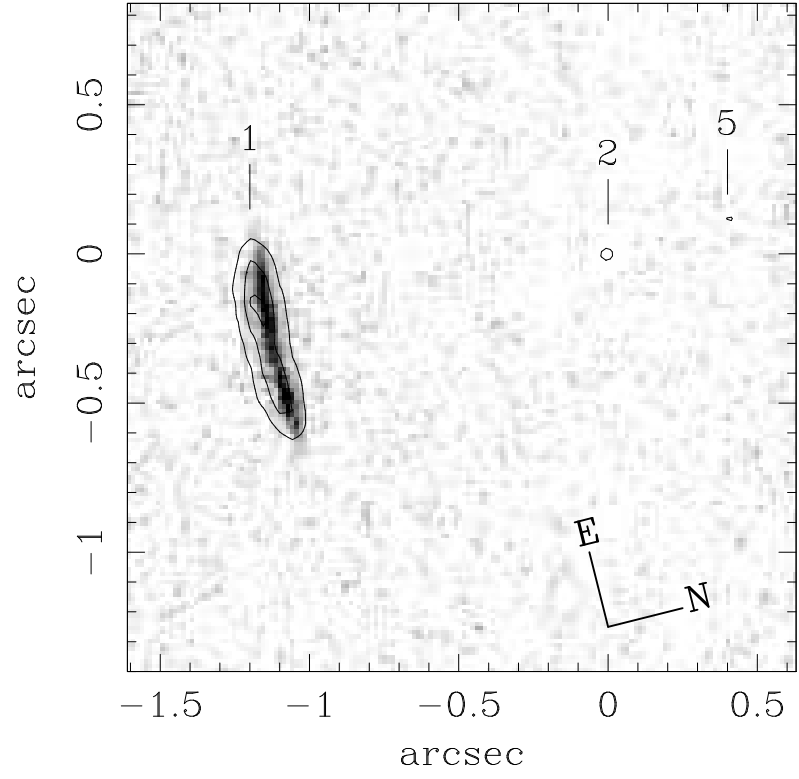
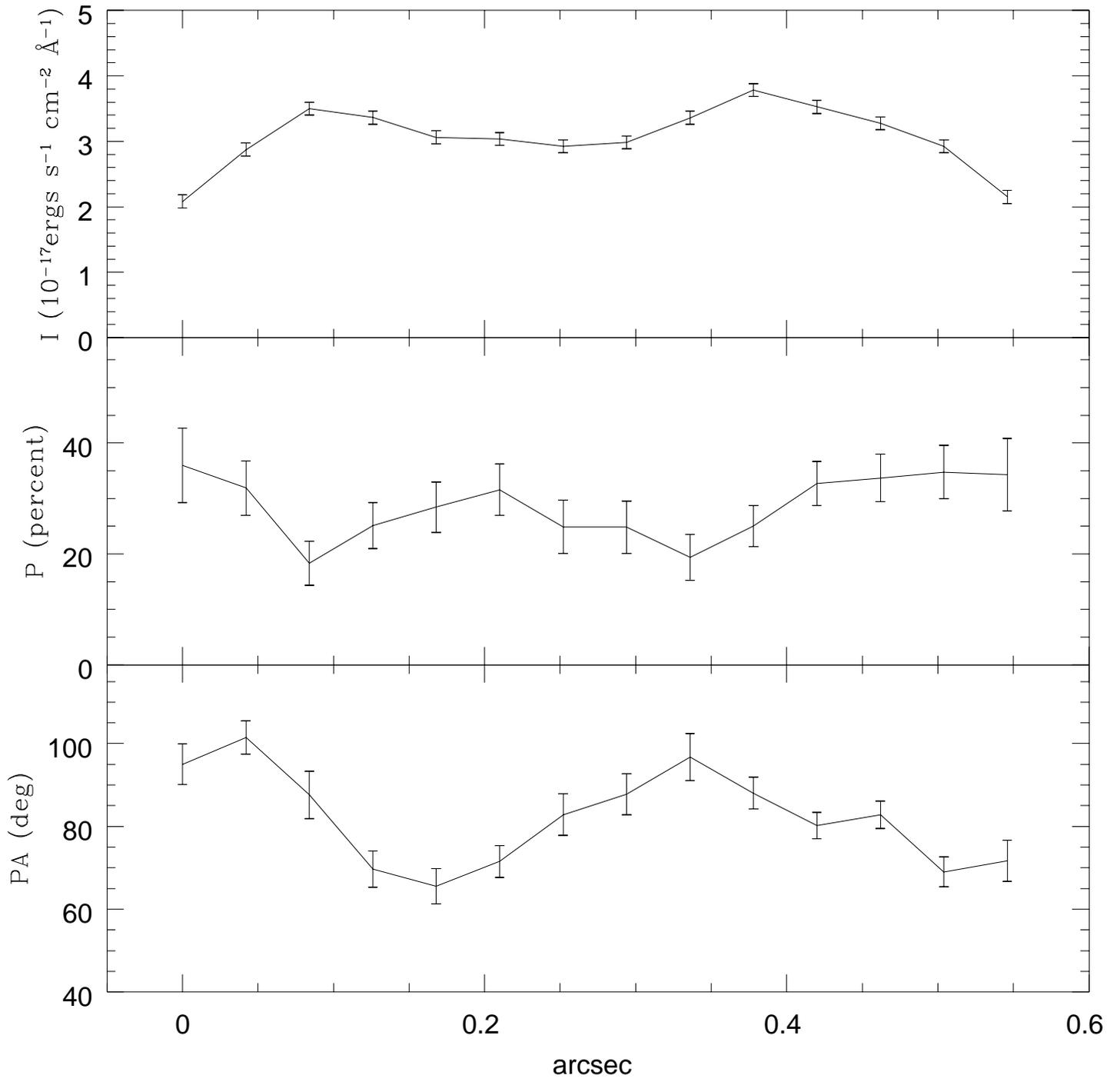
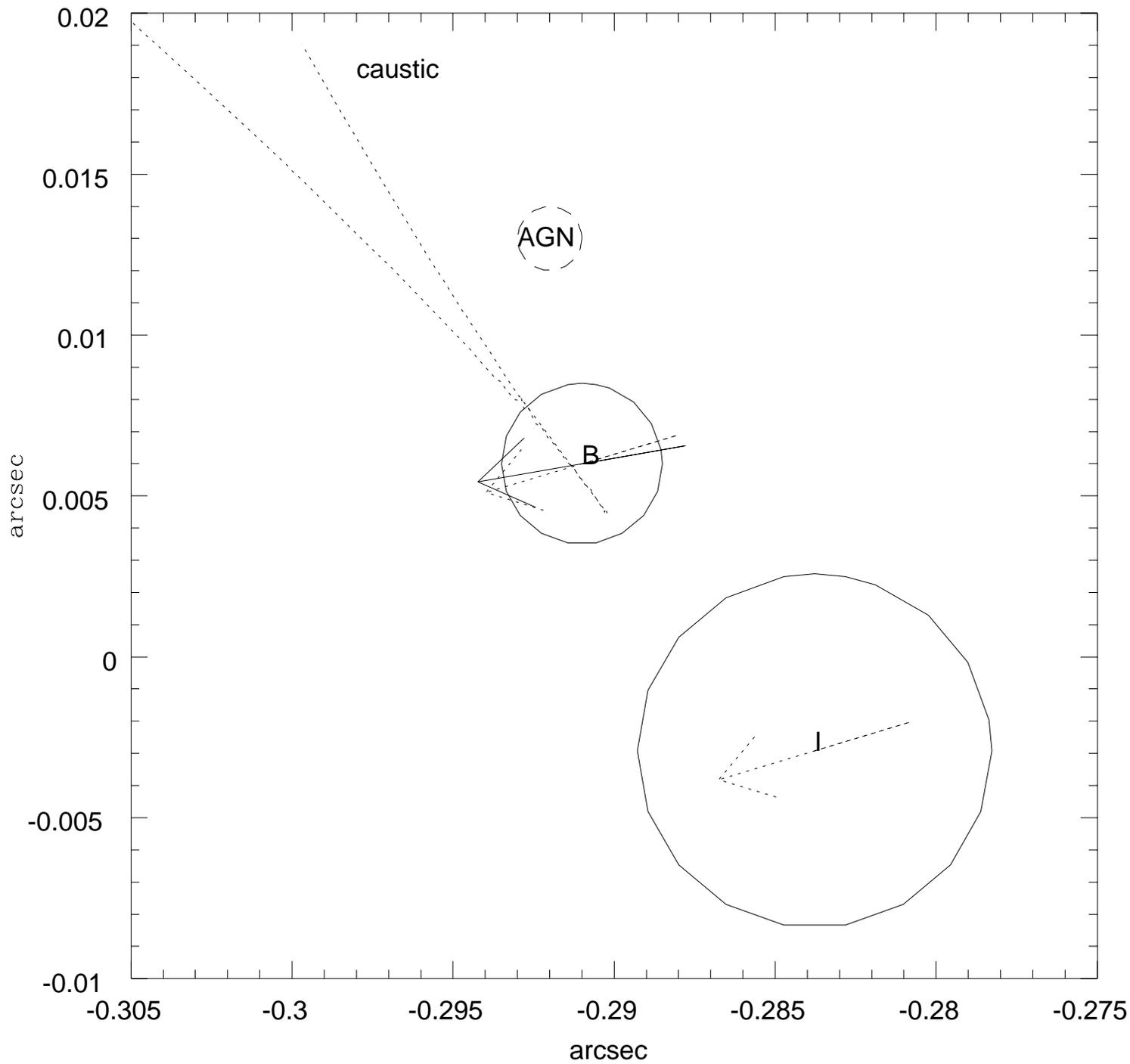
\end{document}